\documentclass[journal]{IEEEtran}

\usepackage[latin9]{inputenc}
\usepackage{textcomp}
\usepackage{amsmath}
\usepackage{amssymb}
\usepackage{tabularx,ragged2e,booktabs}
\usepackage{epsfig,rotating,setspace,latexsym,amsmath,epsf,amssymb,bm}
\usepackage{cite,graphicx,color,subfigure, amsthm}
\usepackage{subfigure}
\usepackage{multicol}
\usepackage{thmtools}
\usepackage{etoolbox}
\usepackage{enumitem}
\usepackage[autostyle]{csquotes}

\newtheorem{prop}{Proposition}

\newtheorem{lemma}{Lemma}

\usepackage{flushend}
\usepackage{lipsum}

\usepackage[english]{babel}

\ifCLASSINFOpdf

\else

\fi

\begin{document}

\title{On the Achievability of Coded Multicasting Fronthaul Delivery in Fog-Aided Networks}

\author{ Seyyed Mohammadreza Azimi
\thanks{S.M.Azimi is with the NJIT,. E-mail: sa677@njit.edu }%
}

\markboth{Draft}%
{Shell \MakeLowercase{\textit{et al.}}: Bare Demo of IEEEtran.cls for IEEE Journals}

\maketitle

\begin{abstract}
Fog  networks benefit from content caching  at edge nodes (ENs)  as well as  fronthaul connectivity to the cloud. In previous works, both  dedicated and multicast fronthaul links has been considered under different assumptions with Normalized  Delivery  Time  (NDT) as a standard to measure the performance. While dedicated links has been studied extensively in different caching scenarios, multicast links has been developed under assumption of $2 \times 2$ fog network where cloud is connected to two ENs serving two users. The extension to arbitrary number of ENs and users only provides the lower bound on the achievable performance. In contrast, this letter proposes an achievable coded multicasting scheme for a general  $M \times K$ fog network with $M$ ENs and $K$ users. The optimality of the proposed scheme is evaluated using both comparison with theoretical lower bound as well as using numerical analysis. \end{abstract}


\IEEEpeerreviewmaketitle

\section{Introduction}
Fog networks paves the way of delivering delay-sensitive content in emerging cellular networks by leveraging local storage at edge nodes (ENs) as well as fronthaul connection to the cloud. Phrasing the inverse-of-degrees of freedom (DoF) as the \textit{Normalized  Delivery  Time  (NDT)} and assuming dedicated fronthaul links between cloud and each EN, lower and upper bounds on the  NDT are obtained in \cite{JSACAzimi}. On the contrary, bounds on the NDT of a $2 \times 2$ fog network with  multicast fronthaul links  is studied in \cite{Koh}. The model is then extended to online edge caching scenario in which the popular files are time-varying \cite{LetterAzimi}.    

\emph{Main Contributions}: 
This letter focuses on fog networks with multicast fronthaul links connecting cloud to more than one EN. The main contributions of this letter are as follows.     

\noindent $\bullet$ Proposing an achievable scheme based on coded multicasting of requested content in fog networks comprised of arbitrary number of users and ENs as opposed to \cite{Koh} in which a 2-by-2 case is considered.  

\noindent $\bullet$ Deriving NDT for the proposed achievable scheme and providing a proof on the order optimality of the scheme.   
 

\section{System Model}\label{sysmo1}
An $M \times K$ fog network is considered in which wireless broadcast fronthaul links connect cloud to $M$ cache-aided ENs serving $K$ users through a downlink wireless channel. Let $\mathcal{F}=\{ F_1,...,F_N\}$ denotes the set of $N$ popular files of size $L$ bits. Defining \emph{fractional cache capacity} as $\mu$ with $0 \leq \mu \leq 1$, each EN can cache $\mu NL$ bits of popular set $\mathcal{F}$ of dimension  $NL$ bits. 
At a given time slot, each user requests a file from the popular set $\mathcal{F}$.  

In any channel use $t \in \{1,...,T\}$, the  received signal at $k$th user is
\begin{equation}\label{rcvr}
Y_{k}(t)=\sum_{m=1}^{M}H_{k,m}X_{m}(t) + Z_{k}(t),
\end{equation}
where $H_{k,m}$ is the channel gain between $m$th EN and $k$th user; $X_{m}(t)$ is the transmitted signal by the $m$th EN; and $Z_{k}(t) \sim \mathcal{CN}(0,1)$ is additive noise at $k$th user. It is assumed that the channel coefficients are independent and identically distributed (i.i.d.) following a continuous distribution. Channel state information (CSI) about the wireless downlink channels, $H=\{\{H_{k,m} \}_{k=1}^{K} \}_{m=1}^{M}$ is available to all the ENs, cloud and users.
At each channel use $t$, the signal received at $m$th EN is 
\begin{equation}\label{rcvr2}
V_{m}(t)=G_{m}U(t)+ W_{m}(t),
\end{equation}
where $G_{m}$ is the wireless channel between cloud and $m$th EN, $U(t)$ denotes the transmitted signal by the cloud in channel use $t$ and $W_{m}(t)$ is additive noise at $m$th EN. The cloud has a power constraint $T^{-1}\sum_{t=1}^{T} \lvert U(t) \rvert ^2 \leq P ^r $ with $T$ is the delivery latency (in channel uses) and $ r \geq 0$ describes the power scaling of the fronthaul transmission as compared to edge transmission. 
CSI about the wireless fronthaul channel $G=\{G_{m} \}_{m=1}^{M}$  is available at all the ENs and cloud. 

 \noindent   \textit{Fronthaul policy}: It is defined by $f_F: \{ d,G,H,\mathcal{F} \} \rightarrow U^T $ with $U^{T}=( U(1),...,U(T) )$ the transmitted message by cloud to the ENs as a function of the demand vector $d=(d_1,...,d_M)$, fronthaul CSI $G$, edge CSI $H$ and the set $\mathcal{F}$. 

 \noindent  \textit{Caching policy}:  It is defined by  $f_C: \mathcal{F}  \rightarrow \{ S_1,...,S_M\}$ in which $S_{m}$ is the cached content at EN $m$. To meet the cache capacity constraint, we have $H(S_{m}) \leq \mu NL$. 

 \noindent \textit{Edge transmission policy}: At EN $m$ and in channel use $t$, it is defined by  $f_E^t: (d,H,G,S_{m},U^{t-1}) \rightarrow X_{m}(t)$ in which transmitted signal at time $t$, namely $X_{m}(t)$, is a function of  demand vector $d$, downlink CSI $H$, fronthaul CSI $G$, cache content $S_{m}$ and the fronthaul messages received at previous channel uses, namely $U^{t-1}= \{U(1),...,U(t-1)\}$.

\noindent \textit{Decoding policy}: It is defined by $f^k_d: \{ d_{k},H,Y_{k} \} \rightarrow \hat{F}_{d_{k}}$ in which user $k$ maps the received signal $Y_{k}$ and downlink CSI $H$ to an estimate $\hat{F}_{d_{k}}$ of the requested file $F_{d_{k}}$.
Denoting a joint fronthaul/caching/edge transmission and decoding policy as $\Pi=(f_C,f_F,f_E^t,f^k_d)$, a sequence of policies $\Pi$ indexed by the file size $L$ is said to be \textit{feasible} if the probability of error 
$\textrm{P}_{e,t}=\underset{k \in \{1,...,K\}}{\max}\textrm{Pr}(\hat{F}_{d_{t,k}} \neq F_{d_{t,k}}) \rightarrow 0 $ when $L \rightarrow \infty$. 

\noindent \textit{Normalized Delivery Time (NDT)}:
For given sequence of feasible policies $\Pi$, the NDT is  defined as \cite{JSACAzimi} 
\begin{equation}\label{NDT_P}
\delta(\mu,r)=\underset{L \rightarrow \infty}{\lim} \underset{P \rightarrow \infty}{\lim}\frac{\textrm{E}_{\mathcal{F},H,d}[T]}{L/\log(P)}.
\end{equation}
$\delta^*(\mu,r)$ denotes the minimum NDT over all feasible policies. Let $T_F$ and $T_E$ be the fronthaul and edge transmission, the corresponding fronthaul and edge NDTs are defined as 
\begin{equation}\label{coop_zf}
\delta_{\text{F}}=\underset{L \rightarrow \infty}{\lim} \underset{P \rightarrow \infty}{\lim}\frac{\textrm{E}_{\mathcal{F},H,d}[T_F]}{L/\log(P)}  ~\text{and}~ \delta_{\text{E}}=\underset{L \rightarrow \infty}{\lim} \underset{P \rightarrow \infty}{\lim}\frac{\textrm{E}_{\mathcal{F},H,d}[T_E]}{L/\log(P)}.
\end{equation}
\section{Delivery Policies}\label{OFF}
This section starts with an overview of cache-aided and cloud-aided delivery, then an achievable scheme using coded multicasting is proposed and finally pipelined transmission is developed using a hybrid of all the different schemes.      
 
 \subsection{Cache-Aided Delivery}\label{CA_Schemes}
 
\noindent $\bullet$ \textit{Cache-Aided Zero Forcing (ZF):} For $\mu=1$,  the set $\mathcal{F}$ is available at each EN and the requested files are delivered using zero forcing beamforming at ENs. The resulting NDTs are \cite{JSACAzimi}: 
\begin{equation}\label{coop_zf}
\delta_{\text{F,ZF}}=0  ~\text{and}~ \delta_{\text{E,ZF}}=\frac{K}{\min\{M,K\}}.
\end{equation}
\noindent $\bullet$ \textit{Cache-Aided Interference Alignment (IA):} If $\mu=1/M$ and each EN has a non-overlapping fraction of set $\mathcal{F}$ such that the entire set is distributed across the ENs, interference alignment is used to deliver the files. The resulting NDTs are \cite{JSACAzimi}: 
\begin{equation}\label{coop_ia}
\delta_{\text{F,IA}}=0  ~\text{and}~ \delta_{\text{E,IA}}=\frac{M+K-1}{M}.
\end{equation}  
 \subsection{Cloud-Aided Delivery}
In this scheme, the requested contents are delivered by using cloud exclusively.  Assuming  the  worst  case in  which  users request $K$ distinct  files, multicasting $KL$ bits on the fronthaul link results in a fronthaul delay of $T_F=KL/(rlog(P))$.  ZF beamforming  is then used for transmission on the edge channel. The resulting edge/fronthaul-NDTs are \cite{JSACAzimi}: 
\begin{equation}\label{cran_e}
\delta_{\text{F,CA}}=\frac{K}{r} ~\text{and} ~ \delta_{\text{E,CA}}=\frac{K}{\min\{M,K\}}.
\end{equation}

\subsection{Coded Multicasting Delivery}
Instead of aforementioned schemes with capacity partitioning, coded multicasting relies on transmission of coded files by treating cached contents at ENs as side information. While Koh et. al  \cite{Koh} focused on  a $2 \times 2$ fog network, this work considers a general $M \times K$ case. The following lemma provides the long-term NDT of coded multicasting scheme. 
\begin{lemma}\label{CC}
\noindent For an $M \times K$ F-RAN with $\mu=1/M$, coded multicasting on fronthaul links results in edge/fronthaul-NDTs as
\begin{equation}\label{cc_ef}
\delta_{\text{F,CC}}=\frac{K(M-1)}{Mr} ~\text{and} ~ \delta_{\text{E,CC}}=\frac{K}{\min\{M,K\}}.
\end{equation}
\end{lemma}      
\begin{proof}
See Appendix \ref{Lemma_1}
\end{proof}

\subsection{Pipelined Transmission Delivery}
 While all the previous schemes relies on serial transmission of files on the fronthaul and edge links, pipelined transmission leverage simultaneous transmission on the fronthaul and edge links. As discussed in \cite{Koh}, the resulting NDT will be 
\begin{equation}\label{def_pipe_off}
\delta_{pl,i}(\mu,r)=\max \big ( \delta_{E,i},\delta_{F,i} \big ),
\end{equation}
with $i \in \{CA, CC, ZF, IA\}$. 
Hence, pipelined transmission with NDT of \eqref{def_pipe_off}  outperforms the serial transmission with an NDT of $\delta_{E,i}+\delta_{F,i}$.  The following lemma provides the NDT of pipelined transmission using cache-aided and coded multicasting schemes.     
\begin{lemma}\label{CCIA}
\noindent For an $M \times K$ F-RAN with $\mu=1/M$, the NDT of pipelined transmission is 
\begin{equation}\label{ccia_ef}
\delta_{\text{pl}}(1/M,r)=
    \left \lbrace \begin{aligned}
      &\delta_{\text{E,IA}}  ~\text{if~} r \leq  r_1\triangleq\frac{K(M-1)}{M+K-1} \\
      &\delta_{\text{F,CC}}  ~\text{if~} r_1 \leq r \leq  r_2\triangleq\frac{(M-1)\min\{M,K\}}{M} \\
      &\delta_{\text{E,CC}}  ~\text{if~} r_2 \leq r \leq  r_3\triangleq \min\{M,K\} \\
      &\delta_{\text{F,CA}}  ~\text{if~} r \geq  r_3
    \end{aligned}
\right.      
\end{equation}

with $\delta_{\text{E,IA}}$ given in \eqref{coop_ia}, $\delta_{\text{F,CC}}, \delta_{\text{E,CC}}$ given in \eqref{cc_ef} and $\delta_{\text{F,CA}}$ given in \eqref{cran_e}.
\end{lemma}      
\begin{proof}
See Appendix \ref{Lemma_2}
\end{proof}
\section{Achievable Scheme}
In this section, an achievable scheme is proposed using different delivery policies discussed in the previous section. 

Let $f_1$ and $f_2$ be two policies with NDTs of $\delta_1$ and $\delta_2$ and cache sizes of $\mu_1$ and $\mu_2$. New policy $f$ with cache size of $\mu=\alpha \mu_1 + (1-\alpha) \mu_2$ and $0 \leq \alpha \leq 1$ is defined by using $f_1$ for $\alpha$ percentage of time and $f_2$ for the remaining time. Hence, the NDT of time sharing-based policy $f$ is \cite{Ali3}
\begin{equation}\label{TS_delta}
\delta= \alpha \delta_1 + (1-\alpha) \delta_2.
\end{equation} 
\begin{prop}\label{ach_offf}
For an $M \times K$ F-RAN with $N \geq K $ and pipelined transmission, achievable NDT is 

\noindent $\bullet$ Low fronthaul $r \leq r_1$
\begin{equation}\label{case_1}
\delta_{\text{ach}}(\mu,r)= \left \lbrace \begin{aligned}
& \mu (M+K-1) + \frac{K(1-\mu M)}{r}  ~~~\text{if~} \mu \leq  \frac{1}{M}  \\
&\frac{1}{M-1} \Bigg ( M+K-1-\mu(\min\{M,K\}-1) \\
&~~~~~~~~~-\frac{K}{\min\{M,K\}} \Bigg)    ~~~~~~\text{if~} \frac{1}{M} \leq \mu \leq 1 \\
\end{aligned}
\right.
\end{equation}
\begin{figure}[t]
\centering
\includegraphics[width=.55\textwidth]{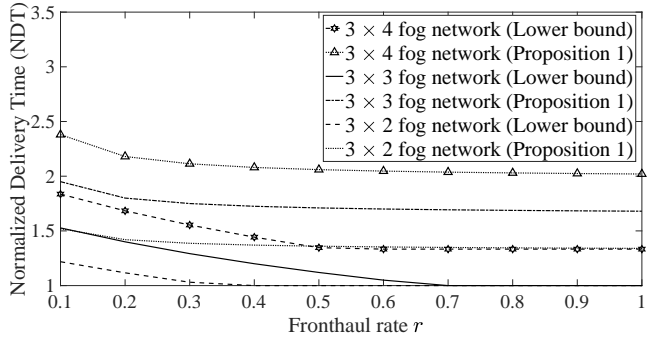}
\caption{NDT as a function of fronthaul rate $r$ for fog networks with $M=3$ edge nodes, $K \in \{2,3,4\}$ end users and $\mu=1/3$.}
\label{ref_on_r}
\vspace{-6mm}
\end{figure}
\noindent $\bullet$ Intermediate fronthaul $I$: $r_1 \leq r \leq r_2$
\begin{equation}\label{case_2}
\delta_{\text{ach}}(\mu,r)= \left \lbrace \begin{aligned}
&\frac{K(1-\mu)}{r}  ~~~~~~~~~~~~~~~~~~~~~\text{if~} \mu \leq  \frac{1}{M} \\
&\frac{K(\mu M-1)}{(M-1)\min\{M,K\}} \\ 
&~~~~~~~~~+\frac{K(1-\mu)}{r}      ~~~~~~~~\text{if~} \frac{1}{M} \leq \mu \leq 1 \\
\end{aligned}
\right.
\end{equation}
\noindent $\bullet$ Intermediate fronthaul $II$: $r_2 \leq r \leq r_3$
\begin{equation}\label{case_3}
\delta_{\text{ach}}(\mu,r)= \left \lbrace \begin{aligned}
&\mu \max\{M,K\} + \frac{K(1-\mu M)}{r} & \text{if~} \mu \leq  \frac{1}{M} \\
&\frac{K}{\min\{M,K\}}    &  \text{if~} \frac{1}{M} \leq \mu \leq 1 \\
\end{aligned}
\right.
\end{equation}
\noindent $\bullet$ High fronthaul $ r \geq r_3$
\begin{equation}\label{case_4}
\delta_{\text{ach}}(\mu,r)=\frac{K}{\min\{M,K\}}         
\end{equation}
with $r_1, r_2$ defined in Lemma \ref{CCIA} and $r_3 \triangleq \min\{M,K\}$. 
\end{prop}
\begin{proof} 
See Appendix \ref{Prop_1} 
\end{proof}

From Proposition \ref{ach_offf}, it can be inferred that for small cache size, i.e. $\mu \leq 1/M$, a fraction of requested files can be delivered only using fronthaul resources and this is due to fact that the corresponding fraction of file is only available in the cloud. Alternatively, when the cache size is large, i.e. $\mu \geq 1/M$, the delivery depends on the fronthaul rate $r$. For large values of $r$, cloud-aided delivery is used since it results in the minimum NDT while for small values of $r$, schemes with edge caching such as cache aided IA or ZF are used in order to achieve the minimum NDT.  The next proposition explains the optimality of proposed achievable scheme.    

\begin{figure}[t]
\centering
\includegraphics[width=.55\textwidth]{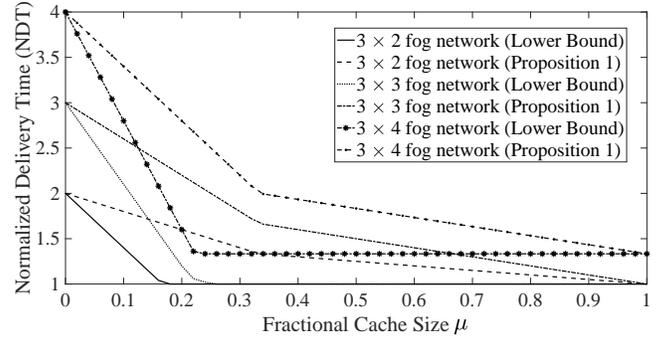}
\caption{NDT as a function of fractional cache size $\mu$ for fog networks with $M=2$ edge nodes, $K \in \{2,3,4\}$ end users and $r=1$.}
\label{ref_on_m}
\vspace{-6mm}
\end{figure}

\begin{prop}\label{ach_order}
For an $M \times K$ F-RAN with $N \geq K $, the achievable NDT in Proposition \ref{ach_offf} satisfies 
\begin{equation}
\delta_{\text{ach}}(\mu,r) \leq 3 \delta^{*}(\mu,r)
\end{equation} 
with $\delta^{*}(\mu,r)$ as the minimum NDT. 
\end{prop}
\begin{proof}
Appendix \ref{Prop_2}
\end{proof}
From Proposition \ref{ach_order}, it can be inferred that the proposed achievable scheme is order optimal with a factor of $3$. Figure \ref{ref_on_r} shows the NDT in Proposition \ref{ach_offf} and lower bound on the NDT \cite{LetterAzimi} as a function of fronthaul rate $r$. For small values of $r$, the NDTs are monotonically decreasing with $r$ while for larger values, the NDTs are fixed due to the fact that only edge transmission has effect on the NDT and it's a function of $M,K$ and $\mu$. For fixed values of $M$ and $\mu$, increasing $K$ reflects the fact that more files will be requested and hence the NDT will be higher.  Figure \ref{ref_on_m} depicts the NDT of proposed scheme and lower bound on the NDT as a function of fractional cache size $\mu$. For small cache size, the NDTs are a decreasing function of $\mu$ while after a certain threshold, NDT is fixed and it is determined by edge transmission limitations. 

\section{Conclusions} \label{Conc}
In this letter, content delivery  in  fog networks  with  wireless broadcast fronthaul links and  edge  connection  is  considered.  To exploit cached contents at ENs as well as the wireless broadcast fronthaul connectivity, an achievable scheme is proposed that relies on the coded multicasting of the requested contents. The NDT is computed as a function of fronthaul rate and fractional cache size and then information-theoretic  analysis is performed to prove the optimality of achievable scheme compared to the minimum theoretical bound. Extension to the scenarios without CSI or with delayed CSI will be considered in the future works.

\appendices
\section{Proof of Lemma \ref{CC}}\label{Lemma_1}

\textit{Caching Policy}: With $\mu=1/M$, each file $F_i \in \mathcal{F}$ is divided into $M$ non-overlapping fractions such that $\cup_{j=1}^{M}F_{i,j}=F_i $ and $\cap_{j=1}^{M}F_{i,j}=\emptyset$. The subfile $F_{i,j}$ is cached at $j^{th}$ EN. 

\textit{Coded Multicasting Policy}: Assuming the worst case in which $K$ users request $K$ different files, without loss of generality it is assumed that files $\{1,..,K\}$ are requested by users $\{1,..,K\}$ respectively.  Cloud transmits the coded files
\begin{equation}\label{cc_encode}
\cup_{i=1}^{K}\cup_{j=1}^{M-1} \{F_{i,j} \oplus F_{i,j+1}\}. 
\end{equation}

\textit{Decoding Policy}: At $j^{th}$ EN, the cached content $F_{i,j}$ and the received files \eqref{cc_encode} are utilized as follows       
\begin{equation}\label{cc_decode}
\cup_{i=1}^{K}F_{i,j} \oplus \big [ \cup_{j=1}^{M-1} \{ F_{i,j} \oplus F_{i,j+1} \} \big ] = \cup_{i=1}^{K}\cup_{j=1}^{M-1} F_{i,j+1}.
\end{equation}
As a results the entire set of popular files $\mathcal{F}$ will be available at every EN. 
Using \eqref{cc_encode}, $K(M-1)$ files, each of them of size $1/M$,  are sent on the fronthaul link and the resulting fronthaul-NDT will be $\delta_{\text{F,CC}}=K(M-1)/(Mr)$.      Since the set of popular files $\mathcal{F}$ is available at every EN, as discussed in Sec. \ref{CA_Schemes}, cache-aided ZF can be used and the resulting fronthaul-NDT will be $\delta_{\text{E,CC}}=K/\min\{M,K\}$. 

\section{Proof of Lemma \ref{CCIA}}\label{Lemma_2}
The NDT of pipelined transmission with cache aided IA is obtained by plugging \eqref{coop_ia} into \eqref{def_pipe_off}. Plugging \eqref{cran_e} into \eqref{def_pipe_off} gives the NDT of pipelined transmission with cloud aided delivery. Similarly, the NDT of pipelined transmission with coded multicasting is obtained by plugging \eqref{cc_ef} into \eqref{def_pipe_off}. If $\delta_{\text{E,IA}} <= \delta_{\text{F,CC}}$ or equivalently $r \leq  r_1\triangleq K(M-1)/(M+K-1)$, the pipelined scheme will be defined by the minimum between two schemes, i.e. cache aided IA. If $\delta_{\text{E,CC}} \leq \delta_{\text{F,CC}} \leq \delta_{\text{E,IA}}$, or equivalently $r_1 \leq r \leq  r_2\triangleq (M-1)\min\{M,K\}/M$, the pipelined NDT of coded multicasting will be determined by its fronthaul NDT. If  $\delta_{\text{F,CC}} \leq \delta_{\text{E,CC}} \leq \delta_{\text{F,CA}} $, or equivalently $r_2 \leq r \leq  r_3 \triangleq \min\{M,K\}$, the pipelined NDT of coded multicasting will be determined by its edge NDT. Finally, if   $\delta_{\text{E,CC}} \geq \delta_{\text{F,CA}} $ or equivalently $ r \geq  r_3$, the pipelined scheme will be using cloud-aided approach. 

\section{Proof of Proposition \ref{ach_offf}}\label{Prop_1}
For $r \leq r_1$ and $\mu \leq 1/M$, using \eqref{def_pipe_off} with $\delta_1=\delta_{pl,IA}(\mu,r)$, $\delta_2=\delta_{pl,CA}(\mu,r)$, $\alpha=\mu M$, plugging $\delta_1, \delta_2$ and $\alpha$  into \eqref{TS_delta} results in the first term in \eqref{case_1}. For $r \leq r_1$ and $\mu \geq 1/M$, using \eqref{def_pipe_off} with $\delta_1=\delta_{pl,IA}(\mu,r)$,  $\delta_2=\delta_{pl,ZF}(\mu,r)$, $\alpha=M(1-\mu)/(M-1)$, plugging $\delta_1, \delta_2$ and $\alpha$  into \eqref{TS_delta} results in the second term in \eqref{case_1}.

For $r_1 \leq r \leq r_2$ and $\mu \leq 1/M$, using \eqref{def_pipe_off} with $\delta_1=\delta_{pl,CC}(\mu,r)$, $\delta_2=\delta_{pl,CA}(\mu,r)$, $\alpha=\mu M$,  plugging $\delta_1, \delta_2$ and $\alpha$  into \eqref{TS_delta} results in the first term in \eqref{case_2}. For $r_1 \leq r \leq r_2$ and $\mu \geq 1/M$, using \eqref{def_pipe_off} with $\delta_1=\delta_{pl,CC}(\mu,r)$,  $\delta_2=\delta_{pl,ZF}(\mu,r)$, $\alpha=M(1-\mu)/(M-1)$,  plugging $\delta_1, \delta_2$ and $\alpha$  into \eqref{TS_delta} results in the second term in \eqref{case_2}.

For $r_2\leq r \leq r_3$ and $\mu \leq 1/M$, using \eqref{def_pipe_off} with $\delta_1=\delta_{pl,CC}(\mu,r)$, $\delta_2=\delta_{pl,CA}(\mu,r)$, $\alpha=\mu M$,  plugging $\delta_1, \delta_2$ and $\alpha$  into \eqref{TS_delta} results in the first term in \eqref{case_3}. For $r_2 \leq r \leq r_3$ and $\mu \geq 1/M$, using \eqref{def_pipe_off} with $\delta_1=\delta_{pl,CC}(\mu,r)$,  $\delta_2=\delta_{pl,ZF}(\mu,r)$, $\alpha=M(1-\mu)/(M-1)$,  and plugging $\delta_1, \delta_2$ and $\alpha$  into \eqref{TS_delta} results in the second term in \eqref{case_3}. 

Finally, for $r \geq r_3$, only cloud aided delivery with pipelined transmission is used. Hence, the achievable NDT is obtained using \eqref{def_pipe_off} with $\delta_{pl,CA}(\mu,r)$.

\section{Proof of Proposition \ref{ach_order}}\label{Prop_2}
First, lower bounds on the minimum NDT is determined. Cut-set bound argument, namely finding a subset of resources which is sufficient to decode a file without error, is used. The resources are the output of fronthaul multicast channel, the output of wireless edge channel and cached contents at ENs. The detailed proof is provided in \cite{LetterAzimi}. The resulting lower bounds on minimum NDT are : 
\begin{equation} \label{Lower_Bounds_1}
\hspace{-15mm} \delta^{*}(\mu,r) \geq \frac{K}{\min \{M,K\}}  
\end{equation}
\begin{equation}\label{Lower_Bounds_2}
\delta^{*}(\mu,r) \geq \frac{K-(K-l)(M-l)\mu}{l+r}, 
\end{equation}
 with $0 \leq l \leq \min \{M,K\}$. Then, the following regimes are considered:

\noindent $\bullet$ Low fronthaul ($r \leq r_1$):

1- Small cache size ($\mu \leq 1/M$): Setting $l=0$ in \eqref{Lower_Bounds_2}, multiplying \eqref{Lower_Bounds_1} by $2$  and subtracting the sum of them from the first term in \eqref{case_1} results in 
\begin{equation}\label{app_1}
\delta_{\text{ach}}(\mu,r)- 3\delta^{*}(\mu,r)= \mu (M+K-1) - \frac{2K}{\min \{M,K\}} \leq 0.
\end{equation} 

2- Large cache size ($1/M \leq \mu \leq 1$): Multiplying \eqref{Lower_Bounds_1} by $2$ and then subtracting the result from the first term in \eqref{case_1} results in
\begin{equation}\label{app_2}
\delta_{\text{ach}}(\mu,r)- 2\delta^{*}(\mu,r)= \mu (M+K-1) - \frac{2K}{\min \{M,K\}} \leq 0.
\end{equation}

\noindent $\bullet$ Intermediate fronthaul $I$: $r_1 \leq r \leq r_2$

1- Small cache size ($\mu \leq 1/M$): Setting $l=0$ in \eqref{Lower_Bounds_2}, multiplying the result by $2$ and then subtracting the result from the first term in \eqref{case_2} results in 
\begin{equation}\label{app_3}
\delta_{\text{ach}}(\mu,r)- 2\delta^{*}(\mu,r)= \frac{K(1-\mu)}{r} - \frac{2K(1-\mu M)}{r} \leq 0.
\end{equation}

2- Large cache size ($1/M \leq \mu \leq 1$): Setting $l=0$ in \eqref{Lower_Bounds_2} , multiplying the result by $2$, summing it with  \eqref{Lower_Bounds_1} and then subtracting the result from the first term in \eqref{case_2} results in 
\begin{equation}\label{app_4}
\begin{aligned}
\delta_{\text{ach}}(\mu,r)- 3\delta^{*}(\mu,r) & = \frac{K(\mu M-1)}{(M-1)\min\{M,K\}} - \frac{2K(1-\mu M)}{r}  \\
 & - \frac{K}{\min \{M,K\}} \leq 0.
\end{aligned}
\end{equation}

\noindent $\bullet$ Intermediate fronthaul $II$: $r_2 \leq r \leq r_3$

1- Small cache size ($\mu \leq 1/M$): Setting $l=0$ in \eqref{Lower_Bounds_2}, summing the result with \eqref{Lower_Bounds_1}  and then subtracting the result from the first term in \eqref{case_3} results in 
\begin{equation}\label{app_5}
\delta_{\text{ach}}(\mu,r)- 2\delta^{*}(\mu,r)= \mu \max\{M,K\} - \frac{K}{\min\{M,K\}}  \leq 0.
\end{equation}

2- Large cache size ($1/M \leq \mu \leq 1$): Comparing \eqref{Lower_Bounds_1}  with first term in \eqref{case_3} reveals that 
\begin{equation}\label{app_6}
\delta_{\text{ach}}(\mu,r)=\delta^{*}(\mu,r)= \frac{K}{\min\{M,K\}}
\end{equation}

\noindent $\bullet$ High fronthaul $ r \geq r_3$ 

Comparing \eqref{Lower_Bounds_1}  with first term in \eqref{case_4} reveals that achievable NDT is the minimum NDT. 

Using \eqref{app_1}-\eqref{app_6} completes the proof.  
\ifCLASSOPTIONcaptionsoff
  \newpage
\fi



%
\bibliographystyle{IEEEtran}
\bibliography{IEEEabrv,IEEEexample}

%




\end{document}